\documentstyle[12pt]{article}

\newcommand{\be}{\begin{equation}}
\newcommand{\ee}{\end{equation}}
\newcommand{\ba}{\begin{eqnarray}}
\newcommand{\ea}{\end{eqnarray}}

\begin{document}
\begin{center}
{\bf COHERENT PION RADIATION FROM }  \\
{\bf  NUCLEON ANTINUCLEON ANNIHILATION}
\end{center}
\vspace{0.5cm}
\begin{center}
R.D.~Amado$^{a,b}$, F.~Cannata$^{a,c}$,
J.-P.~Dedonder$^{a,d}$,  M.P.~Locher$^a$ and \\
{}~Bin ~Shao$^b$  \\
$^a$ Paul Scherrer Institute, CH-5232 Villigen-PSI, Switzerland \\
$^b$ Department of Physics, University of Pennsylvania, Philadelphia,
PA 19104, USA  \footnote{Permanent address of RDA} \\
$^c$ Dipartimento di Fisica and INFN,
I-40126 Bologna, Italy \footnote{Permanent address of FC} \\
$^d$ Laboratoire de Physique
Nucl\'eaire, Universit\'e Paris 7, \\
2 Place Jussieu, F-75251 Paris Cedex 05 and
Division de \\ \mbox{}\hspace{0.7cm} Physique Th\'eorique,
IPN, F-91406 Orsay, France  \footnote{Permanent addresses of JPD}
\end{center}

\vspace{0.5cm}
\today
\begin{abstract}
A unified picture of nucleon antinucleon annihilation into
pions emerges from a classical description of the pion
wave produced in annihilation and the subsequent
quantization of that wave as a coherent state.  When
the constraints of energy-momentum and iso-spin conservation
are imposed on the coherent state, the pion
number distribution and charge ratios are
found to be in excellent agreement
with experiment.
\end{abstract}
\newpage

Nucleon antinucleon annihilation at rest or at low energy
goes mostly into pions. The problem for theory is to calculate
the pion spectrum, the average number of pions, the branching
rate into various pion modes, higher moments of the pion
distribution, correlations and charge
ratios, all from some theory of the
annihilation process.  In this note we show that a unified
account of all these features comes
naturally out of a picture
of very rapid annihilation into a classical pion wave
and subsequent quantum description of this wave as a
coherent state.  A very simple parameterization of the
annihilation process coupled with the constraints of iso-spin
and four-momentum conservation then leads to an excellent
account of the principal features of the annihilation spectrum.

Recent studies of annihilation in the Skyrme model \cite{Sommerman}
\cite{SWA}
have suggested that annihilation proceeds very rapidly when
the baryon and anti-baryon collide and that the product of this
rapid annihilation is a pion pulse or coherent pion wave. This
is a classical picture.  We can quantize
that wave by using the method of coherent states \cite{Glauber}.
The relatively large energy released in annihilation
at rest (13.9 pion masses) and the moderately large
number of pions seen on average ($\sim 5$) makes a classical
starting point plausible.  The empirical success of the coherent
state description gives the quantized rapid pion pulse
account credibility.

Suppose the pion wave from annihilation has a source
$S(\vec{r},t)$, that is well localized in space and time,
with a corresponding Fourier transform $s(\vec{k},\omega)$.
If we assume the pion wave obeys a linear wave equation with this
source, the pion radiation field, $\phi(\vec{r},t)$ for
$t>0$ is given by
\be
\phi(\vec{r},t) = -\int \frac{i d^3k}{4 \pi \omega_k}
s(\vec{k},\omega_k) e^{i \vec{k} \cdot \vec{r} -i \omega_k t}
\ee
where $\omega_k=\sqrt{k^2+\mu^2}$ and $\mu$ is the pion mass.
Introducing creation and annihilation operators for each
mode $\vec{k}$, $a_{\vec{k}}^{\dagger}$ and $a_{\vec{k}}$,
one can define the coherent quantum state corresponding to
 $\phi(\vec{r},t)$ by
 \be
 |f \rangle = e^{-\hat{N}/2}
 e^{\int f(\vec{k}) d^3k a_{\vec{k}}^{\dagger}}|0 \rangle
 \ee
 with
 \be
 f(\vec{k})=-i \sqrt{2 \pi} \frac{s(\vec{k},\omega_k)}{2 \omega_k}.
\ee
The mean number of pions in the state $|f\rangle$ is
\be
 \hat{N} =\int |f(\vec{k})|^2 d^3k
 \ee
and$|f(\vec{k})|^2$
is easily seen to be the single pion momentum distribution
in that state.
The mean energy released is
\be
\hat{E} = \int \omega_k |f(\vec{k})|^2 d^3k
\ee
We will set this energy equal to the energy released
in annihilation.

The coherent state contains all possible numbers of
pions and hence treats all pion annihilation channels
together and in a unified way.
These channels are
distributed in a Poisson distribution  with
mean number $\hat{N}$ and variance $\sigma = \sqrt{\hat{N}}$.
We arrange parameters so that $\hat{N} \sim 6$.
We have chosen $6$ rather than the experimental $5$ to
roughly compensate for decay into resonant channels.
These contribute to making $\hat{N}$ small, but are not part of
our ``pions only" picture.
With this $\hat{N}$,
we find too large a variance.  The experimental
result is $\sigma \sim 1$.  Furthermore only states
with between 2 and 13 pions are permitted in annihilation
at rest by energy and momentum conservation, while the
coherent state contains any number of pions.
To maintain the simultaneous treatment of all
annihilation channels that is characteristic of the coherent
states, and correct its failings, we need to introduce
the constraints of
energy and momentum \cite{HornSilver}.
To do this, introduce the operator
$F(x)$,
\be
F(x)= \int d^3k f(\vec{k})a_{\vec{k}}^{\dagger} e^{i k \cdot x}
\ee
that creates a state at the four-vector position $x$ with
Fourier components $ f(\vec{k})$ (the same  $ f(\vec{k})$ as above).
In (6) the time component of $k$ is the on shell energy $\omega_k$.
The state of fixed total four-momentum $K$ is then given by
\be
|f,K \rangle = \int \frac{d^4x}{(2 \pi)^4} e^{-i K \cdot x} e^{F(x)}
|0 \rangle
\ee
as is easily checked by expanding out the exponential  $e^{F(x)}$.
One finds $\langle f,K'|f,K \rangle = \delta^4(K-K')I(K)$, with
\be
 I(K) =\int \frac{d^4x}{(2 \pi)^4} e^{i K \cdot x}
 e^{\rho(x)}
 \ee
 and with
 \be
 \rho(x) = \int d^3p |f(\vec{p})|^2 e^{-i p \cdot x}
 \ee
The integral for $I(K)$ is singular because of the large $x$ behavior
of the factor $e^{\rho(x)}$.  This singular behavior comes from the
``one" in the expansion of this exponent which comes in turn from the
``one" in the expansion of $e^{F(x)}$.  This is the no pion state
and is forbidden by four-momentum conservation,  as is the one pion state.
Hence we can write
\ba
     I(K)&=&\int \frac{d^4x}{(2 \pi)^4} e^{i K \cdot x}
     \sum_{m=2} \frac{\rho^m(x)}{m!}    \nonumber \\
     &=& \sum_{m=2} \frac{I_m(K)}{m!}
\ea
The sum over $m$ terminates with $m=13$ by four-momentum conservation.
The $I_m(K)$ are easily calculated by first doing the $x$ integration,
extracting the four-momentum conserving delta function and calculating
the remaining integrals with the constraint imposed.  In terms of
the $I_m$, the probability of finding $m$ pions
is $P_m= \frac{I_m(K)}{I(K) m!}$
and the mean number of pions is given by
\be
\hat{N}=\sum_{m=2} \frac{m I_m(K)}{I(K) m!}
\ee

To calculate $P_m$ and $\hat{N}$, we need to model $S(\vec{r},t)$.
Inspired by the Skyrmion calculations, \cite{Sommerman}  \cite {SWA}
we take a very simple spherically symmetric form. We assume
$S(\vec{r},t)= 0$ for $t<0$ and for $t>0$,
\be
S(\vec{r},t) = C t e^{-\gamma t} e^{-\alpha r}/r
\ee
where $C$ is a scale constant. This leads to
\be
|f(\vec{k})|^2=\frac{C' k^2}{(k^2+\alpha^2)^2(\omega_k^2+\gamma^2)^2
\omega_k^2}
\ee
where $C'$ is a another scale constant and where we have
multiplied $f$ by $k$ to model the p-wave nature of pion emission.
We fix $C'$ by requiring that the average energy be the energy emitted
in annihilation at rest,
(5), which is $13.87$ in units of the pion mass, ($\mu=1$).
For the range parameters we make the very simple assumption that
$\alpha=\gamma$, and after very little parameter searching find that
$\alpha=\gamma=2$ gives $\hat{N}=6$ for the Poisson distribution.
Thus this very simple form with the reasonable size of half the
pion Compton wave length begins to look like the data.
Using the {\bf same} $f(\vec{k})$ in the four-momentum restricted
calculation we find $\hat{N}=6.4$ and $\sigma=.88$, both in quite
close agreement with experiment roughly corrected for final
state resonances \cite{S&S}, \cite{DoverGuts}.

In Figure 1 we show the probability of finding $m$ pions in annihilation
at rest calculated with the unconstrained form (Poisson distribution),
and with the constraint of four-momentum conservation. We see that
the constraint sharply narrows the distribution. In fact on the scale of
the figure, the constrained distribution is indistinguishable from
a Gaussian distribution of the same average and variance.  The
constrained distribution plotted in Figure 1 is nearly identical to
the distribution seen in experiment \cite{DoverGuts}, \cite{11}.

So far we have neglected the iso-spin of the pion.  Since
for averages, the constraint of four-momentum conservation does
not seem very important, let us examine the effects of iso-spin
on pion averages in the ordinary coherent state.
We must now make the pion creation operator an iso-vector.  But we
cannot make $f(\vec{k})$ an iso-vector and dot it into the
creation operator because iso-spin conservation is a global
constraint. It must hold for every $k$. To
implement iso-spin conservation, we introduce
a fixed unit vector in iso-spin space, $\hat{T}$,
dotted into the creation operator, and write the
coherent state with fixed total iso-spin, $I$, and z-component
$I_z$ as
\be
|f,I,I_z \rangle = \nu \int \frac{d \hat{T}}{\sqrt{4 \pi}} Y^*_{I,I_Z}(
\hat{T}) e^{\frac{-\hat{N}}{2}} e^{\int f(\vec{k}) d^3k
a^{\dagger}_{\vec{k},\mu} \hat{T}_{\mu}} |0\rangle
\ee
where $\nu$ is a normalization factor, and $\hat{N}$ is the average
number of pions summed over charge types. It is still  given by
(4).  The expectation value of the
number operator for pions  of charge type $\mu$
in this state gives the average number of pions of this type.
For $I=0$ one easily finds the expected answer of $\hat{N}_{\mu}
=\hat{N}/3$, for any $\mu$.  For $I=1$, $I_z=0$ on finds,
$\hat{N}_+=\hat{N}_-=\hat{N}/5$ and $\hat{N}_0=3\hat{N}/5$.  These
results are obtained in the large $\hat{N}$ limit where they are
independent of the form of $f(\vec{k})$.  Note the excess
of $\pi_0$ in the $I=1$ case.  If we use a recent estimate of the
relative population of $I=0$ and $I=1$ in proton antiproton annihilation
at rest \cite{ZL}, we find $\hat{N}_0/\hat{N}_+ =1.53 \pm .15$ (the
uncertainty comes from the population estimate), to be compared with
the experimental value of $1.27 \pm .14$ \cite{S&S},\cite{11}.
Thus the  coherent state constrained by iso-spin conservation
agrees remarkably well with the data for the pion charge ratios
and naturally accounts for the excess of $\pi_0$'s.

We have seen that a description of proton antiproton annihilation
at rest into a classical pion wave and the subsequent
quantization of that wave as a coherent state gives a unified picture
of all the direct pion channels and correctly accounts for the
number distributions of those channels and for the charge ratios,
all with very few parameters.  Imposing four-momentum and iso-spin
conservation on the  coherent state is important for this agreement.
These constraints also introduce correlations that are not in the
simple coherent state.  Extensions of these methods to study these
correlations and to other conservation laws are under study.
Further afield one can imagine applying these ideas to any strongly
interacting process in which the original event might be described
classically and quantum mechanics restored by coherent states.
Jets, Centauros, and very high energy
heavy ion collisions may be three such cases.
For Centauros and heavy ions
there has recently been some work along
these lines  \cite{Greiner}, \cite{Blaz}, \cite{Raj}.
These approaches, like ours, take classical, nonperturbative
QCD as their starting point.
Much of this work uses polynomial coherent states with
fixed numbers of pions in order to
introduce quantum mechanics and to implement
iso-spin conservation.  The very large number of pions
($>100$) seen in these processes suggests that field
coherent states with our prescription for iso-spin
are a more natural approach.
One might imagine many other processes in which
QCD can be treated classically and then experimental
manifestations described by coherent states.
In many cases, including annihilation,
it may be necessary to further develop
the treatment either by introducing
squeezed states or by using a density matrix to
average over coherent states  when processes have important
statistical components as in the case of heavy ions.

RDA, FC, and JPD  would like to thank
the theory group of the Division of Nuclear
and Particle Physics at
the Paul Scherrer Institute
for providing a very
pleasant environment in which much of this work was done.  RDA
and BS are
supported in part by the United States National Science Foundation.
The Division de Physique Th\'eorique is a Research Unit of the
Universities Paris 11 and 6 associated to CNRS.

\newpage
\begin{center}
Figure Caption
\end{center}
\begin{flushleft}
Fig.1 The probability, $P_m$, of having $m$ pions in
nucleon antinucleon annihilation at rest as a function
of $m$. The open circles refer to the unconstrained
coherent state and are a Poisson distribution. The
solid squares include the constraint of
four-momentum conservation. This set of points agrees
well with experiment, \cite{DoverGuts}, \cite{11}, and
is indistinguishable (on the scale of the figure)
from a Gaussian distribution of the same mean and variance.
\end{flushleft}

\end{document}